\documentclass[journal,twocolumn]{IEEEtran}

\usepackage{cite}
\usepackage{amsmath,amsfonts,amsxtra,amssymb,latexsym,amscd,amsthm,mathrsfs,bm}
\usepackage{graphicx}
\graphicspath{{./figures/}}
\usepackage{hyperref}
\usepackage{xcolor}
\usepackage{algorithm}
\usepackage{algorithmic}
\usepackage[small]{caption} 
\usepackage{epstopdf}
\usepackage{comment}

\newcommand{\RomanNumeralCaps}[1]
    {\MakeUppercase{\romannumeral #1}}
\newcounter{MYtempeqncnt}

\title{Pilot Design and Doubly-Selective Channel Estimation for Faster-than-Nyquist Signaling}          

\author{Simin Keykhosravi and Ebrahim Bedeer}

\begin{document}
\frenchspacing 

\maketitle
\begin{abstract}
Being capable of enhancing the spectral efficiency (SE), faster-than-Nyquist (FTN) signaling is a promising approach for wireless communication systems. This paper investigates the doubly-selective (i.e., time- and frequency-selective) channel estimation and data detection of FTN signaling. We consider the intersymbol interference (ISI)  resulting from both the FTN signaling and the frequency-selective channel and adopt an efficient frame structure with reduced overhead. We propose a novel channel estimation technique of FTN signaling based on the least sum of squared errors (LSSE) approach to estimate the complex channel coefficients at the pilot locations within the frame. In particular, we find the optimal pilot sequence that minimizes the mean square error (MSE) of the channel estimation. To address the time-selective nature of the channel, we use a low-complexity linear interpolation to track the complex channel coefficients at the data symbols locations within the frame. To detect the data symbols of FTN signaling, we adopt a turbo equalization technique based on a linear soft-input soft-output (SISO) minimum mean square error (MMSE) equalizer. Simulation results show that the MSE of the proposed FTN signaling channel estimation employing the designed optimal pilot sequence is lower than its counterpart designed for conventional Nyquist transmission. The bit error rate (BER) of the FTN signaling employing the proposed optimal pilot sequence shows improvement compared to the FTN signaling employing the conventional Nyquist pilot sequence. Additionally, for the same SE, the proposed FTN signaling channel estimation employing the designed optimal pilot sequence shows better performance when compared to competing techniques from the literature.
\end{abstract}


\section{Introduction}
The tremendous growth in data traffic in recent years and the increasingly scarce bandwidth have placed a heavy demand on improving the spectral efficiency (SE) of wireless communication systems. Introduced in 1970s, faster-than-Nyquist (FTN) signaling has recently redrawn the interests of researchers owing to its capability to enhance the SE without necessarily increasing the bandwidth or the energy per bit \cite{anderson2013faster}. 

In the conventional Nyquist transmission using $T$-orthogonal pulses, the Nyquist limit indicates a maximum rate of $1/T$ to have an inter-symbol interference (ISI)-free communication over frequency flat channels \cite{john2008digital}. However, FTN signaling involves the transmission of data symbols using $T$-orthogonal pulses at every symbol interval $\tau T$, where $0 < \tau \leq 1$ is the FTN signaling acceleration parameter. Intentionally violating the Nyquist limit increases the transmission rate; however, at the price of unavoidable ISI at the receiver. Mazo inspected the minimum square Euclidean distance associated with the transmission of uncoded binary sinc pulses and showed that it remains constant, and thus the asymptotic error rate, when decreasing $\tau$ to a specific limit of $0.802$ \cite{mazo1975faster}---this limit was later known as the Mazo limit. Such acceleration results in $25\%$ times higher rate when compared to the conventional Nyquist transmission employing the same energy per bit and bandwidth.  

Since the optimal detection of FTN signaling for small values of $\tau$, even in additive white Gaussian noise (AWGN), suffers from high computational complexity, several works approximated the optimal solution by using reduced trellis or reduced tree search \cite{liveris2003exploiting, prlja2008receivers, bedeer2017reduced}. These trellis-based equalizers are too complex to be utilized in applications where the number of bits per symbol is high. When trellis-based equalizers cannot be used
due to complexity, filter-based equalizers would be a promising option. In the early works of FTN signaling over AWGN channel, time-domain optimal detectors have been designed to alleviate the ISI \cite{liveris2003exploiting}, \cite{prlja2008receivers}. In mild ISI conditions, i.e., high values of $\tau$, very low complexity detectors, based on frequency domain equalization (FDE) \cite{sugiura2013frequency} or symbol-by-symbol detection \cite{bedeer2017very}, are used to strike a balance between the complexity and detection performance. The detection of FTN signaling becomes more challenging for high-order and ultra-high order modulations, and precoding techniques \cite{li2020beyond} and optimization methods based on alternating directions multiplier method (ADMM) \cite{ibrahim2021novel} are effective to reduce the computational complexity.

A few works considered the detection of FTN signaling over frequency-selective channels \cite{hirano2014tdm, wu2017hybrid, ishihara2017iterative, shi2017frequency}. In particular, in \cite{hirano2014tdm}, a pilot sequence based on the Nyquist criterion was inserted preceding the data block to estimate the channel. Unlike \cite{hirano2014tdm}, a pilot sequence based on FTN signaling criterion was employed in the time-domain joint channel estimation and FTN signaling detection in \cite{wu2017hybrid}. To reduce the complexity of the channel estimation and FTN signaling detection, the work in \cite{ishihara2017iterative} designed a low complexity joint channel estimation and data detection for FTN signaling in the frequency-domain. However, to avoid the ISI from the FTN signaling pilot sequence, a guard block of the same length as the pilot sequence is inserted before the transmitted data in the frame. Additionally, a cyclic prefix of the same length as the pilot sequence and the guard block is added at the end of the transmitted data in the frame.  This results in adding a total overhead of four times the pilot sequence length within the  frame, and eventually, reduced the SE of the design in \cite{ishihara2017iterative}. In contrast to \cite{hirano2014tdm, wu2017hybrid, ishihara2017iterative} where the channel is considered to be a quasi-static, \cite{shi2017frequency} considered a time-selective channel and proposed a frequency-domain joint channel estimation and FTN signaling detection technique based on the message passing algorithm.

This paper contributions can be summarized as: 
\begin{itemize}
  \item This paper scrutinizes the channel estimation and data detection for FTN signaling over doubly-selective (i.e. time- and frequency-selective) channels. We adopt a frame structure considering the ISI introduced by both FTN signaling and the frequency-selective channel. The adopted frame structure does not suffer from SE loss as it does not require the additional overhead of a cyclic prefix. It should be mentioned that the detrimental effects of inter-block interference (IBI) are also considered in the proposed frame structure.

  \item In order to estimate the complex channel coefficients at the pilot locations within the frame, we propose a channel estimation approach based on the least sum of squared errors (LSSE) method. Additionally, we find the optimal FTN signaling pilot sequence that minimizes the mean square error (MSE) of the channel estimation. To resolve the time-selective nature of the channel, we employ a low-complexity linear interpolation to track the channel over the data symbols part of the frame.

  \item
  The proposed approach in this paper is evaluated through extensive simulations for FTN signaling transmission over a doubly-selective Rayleigh fading channel.
  Simulation results show that the MSE of the proposed FTN signaling channel estimation employing the designed optimal pilot sequence is lower than its counterpart designed for conventional Nyquist transmission. The bit error rate (BER) of the FTN signaling employing the proposed optimal pilot sequence shows improvements compared to the FTN signaling employing the conventional Nyquist pilot sequence.
  Simulation results, for the same SE over a frequency-selective Rayleigh fading channel, show a significant improvement (about 6 dB) in the MSE of the proposed FTN signaling channel estimation employing the designed optimal pilot sequence when compared to the technique introduced in \cite{ishihara2017iterative}.
\end{itemize}
The remainder of this paper is organized as follows. Section 2 presents the system model of the FTN signaling in the doubly-selective channel and introduces the adopted frame structure. Section 3 discusses the channel estimation, the optimal design of pilot sequence, and the channel tracking. 
The data detection of FTN signaling using a linear turbo equalizer is discussed in Section 4. Section 5 provides the simulation results. Finally, Section 6 concludes the paper.
\begin{figure*}[t]
\centering
\includegraphics[width=0.75\textwidth]{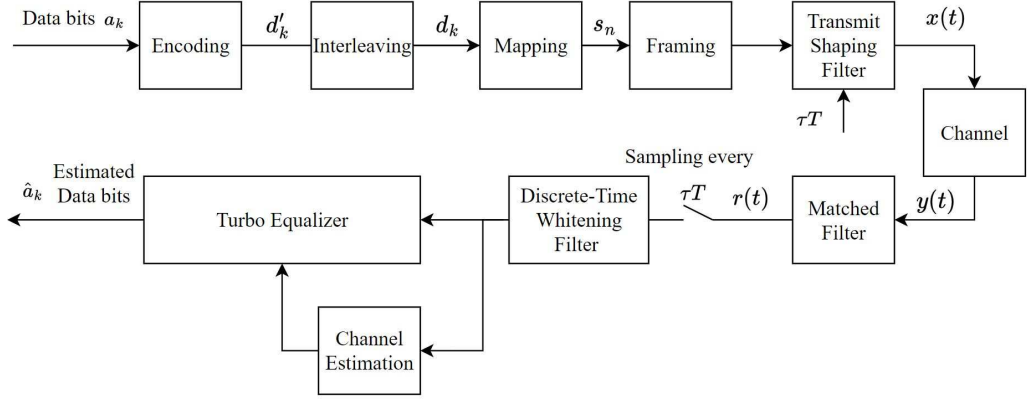}
\caption{Block diagram of a communication system employing FTN signaling.}
\label{fig:1}
\end{figure*}
Throughout the paper, boldface uppercase letters, e.g., $\mathbf{X}$, are used to denote matrices, boldface lowercase letters, e.g. $\mathbf{x}$, are used to denote column vectors, and lightface lowercase letters, e.g., $x$, are used to denote scalers. 
We use $\mathbf{X}^{-1}$, $\mathbf{X}^\text{T}$, and $\mathbf{X}^\text{H}$ to denote the inverse, transpose, complex conjugate transpose (Hermitian transpose) operations of matrix $\mathbf{X}$, respectively.  We use the notation  $x^\ast$ to denote the complex conjugate of a complex number $x$, while the convolution operator between $x$ and $y$ is denoted by $x \ast y$. We use $\text{Pr}(x=a)$ to denote that the probability of a discrete random variable $x$ equals $a$,  $\text{E}(\mathbf{x})$ to represent the expectation operator of $\mathbf{x}$, and $\text{Cov}(\textbf{x}, \textbf{y})$ to denote the covariance of $\textbf{x}$ and $\textbf{y}$. The operator $\text{Diag}(\textbf{x})$ applied to a vector $\textbf{x}$ of length $N$ denotes an $N \times N$ diagonal matrix constructed from the vector $\textbf{x}$. The $\text{tr}({\mathbf{X}})$ denotes the trace operation of matrix $\mathbf{X}$. Finally, $\mathbb{C}^{N\times M}$ includes all complex matrices with dimension ${N\times M}$.

\section{System Model}
The FTN signaling system employed in this paper is depicted in Fig. \ref{fig:1}. On the transmitter side, each block of independent and identically distributed (i.i.d.) data bits $\textbf{a}$ is encoded to  $\textbf{d}'$. To assist in mitigating the effect of error bursts due to the fading channels, the interleaver shuffles each block of coded bits to $\textbf{d}$. Then, each consecutive ${\log_2}M$ output bits of the interleaver are modulated onto a complex data symbol that belongs to an $M$-ary quadrature amplitude modulation ($M$-QAM) constellation set $\mathcal{S}_\text{d}$.

To aid the receiver in the channel estimation, known modulated symbols, i.e., pilots, of length $N_\text{p}$ are inserted every $N_\text{d}$ data symbol block to form frames of length $N = N_\text{p} + N_\text{d}$. Pilots and data symbols in the $i$th frame are denoted as ${s}_n, \, n= (i-1)N+k$, where $i = 1, \ldots, \infty$, is the frame index and $k=0, 1, \ldots, N-1$, is the pilot and data symbols index within a frame. In particular, within a given frame, index $k=0, 1, \ldots, N_\text{p}-1$ denotes the pilot symbols and the index $k=N_\text{p}, 1, \ldots, N-1$ indicates the data symbols. To simplify the notation, we drop the frame index $i$, and denote the symbols as ${s}_n, \, n=0, 1, \ldots, \infty$.  

The pilot and data symbols within a frame are passed through a $T$-orthogonal root raised cosine (RRC) pulse, $g(t)$, having unit energy, i.e., $\int_{-\infty}^{\infty} |g(t)|^2 \,dt = 1$, and then, transmitted every symbol interval $\tau T$. That said, the baseband transmitted signal can be expressed as 
\begin{IEEEeqnarray}{rcl}\label{equation1}
x(t) &{}={}& \sum\limits_{n=0}^{\infty} {s_n g(t-n\tau T)}.
\end{IEEEeqnarray}

As discussed earlier, we consider the transmission of FTN signaling over a doubly-selective fading channel denoted as $c(t,\phi)$; hence, the received signal can be represented as
\begin{IEEEeqnarray}{rcl}\label{equation2}
y(t) &{}={}& x(t)\ast c(t,\phi) + n(t),
\end{IEEEeqnarray}
where $n(t)$ is the zero-mean AWGN with variance $\sigma_n^2$. The received signal, $y(t)$, is passed through a filter matched to $g(t)$, and the signal at the matched filter output can be expressed as
\begin{IEEEeqnarray}{rcl}\label{equation4}
r(t) &{}={}& x(t)\ast  c(t,\phi)\ast g(t)+z(t),
\end{IEEEeqnarray}
where $z(t) = n(t)\ast g(t)$. 
Substituting $s(t)$ from (\ref{equation1}), the received signal in (\ref{equation4}) is re-expressed as
\begin{IEEEeqnarray}{rcl}\label{equation5}
r(t) =  \sum\limits_{n=0}^{\infty} {s_n h(t-n\tau T)}\ast c(t,\phi)+z(t),
\end{IEEEeqnarray}
where $h(t) = g(t)\ast g(t)$. Using the tapped delay line channel model \cite{cavers2006mobile}, a doubly-selective channel with the impulse response of $c(t,\phi)$ is considered to have $L_\text{c}+1$ channel taps, $c_l(t),\, l =0, 1, \ldots, {L_\text{c}}$, with the $l$th channel path $c_l(t)$ has a delay of $\phi_l$.

The matched-filtered signal $r(t)$ is sampled every $\tau T$ and the $k$th received sample can be written as
\begin{IEEEeqnarray}{rcl}\label{equation6}
r_k &{}={}& r(k\tau T) \nonumber\\
&{}={}&\sum\limits_{n=0}^{\infty} {s_n} \sum\limits_{l=0}^{L_\text{c}} c_l(k\tau T) h \left( \left(k-n-l\right) \tau T \right)+z(k\tau T).
\end{IEEEeqnarray}
After a change of variable $n=k-j$, the $k$th received sample in (\ref{equation6}) can be rewritten as
\begin{IEEEeqnarray}{rcl}\label{equation7}
r_k &{}={}& \sum\limits_{l=0}^{L_\text{c}} c_{k,l} \sum\limits_{j=-\infty}^{k} s_{k-j} h_{j-l} +z_k,
\end{IEEEeqnarray}
where $c_{k,l}=c_l(k\tau T)$, $h_{j-l}=h\left(\left(j-l\right)\tau T\right)$, and $z_k=z(k\tau T)$. One can show that the noise samples $z_k$ are colored, and hence, a discrete-time noise whitening filter can be used before channel estimation and FTN signaling detection. 

In the coming subsection, we adopt a frame structure with reduced overhead. The received samples introduced in (\ref{equation7}) then pass through the whitening filter and fed into the channel estimation and turbo equalizer. Turbo equalizer also receives the estimated channel impulse response as an input from the channel estimation block and outputs the estimated data bits $\hat{a}_k$.
\begin{figure*}[t]
\centering
\includegraphics[width=0.75\textwidth]{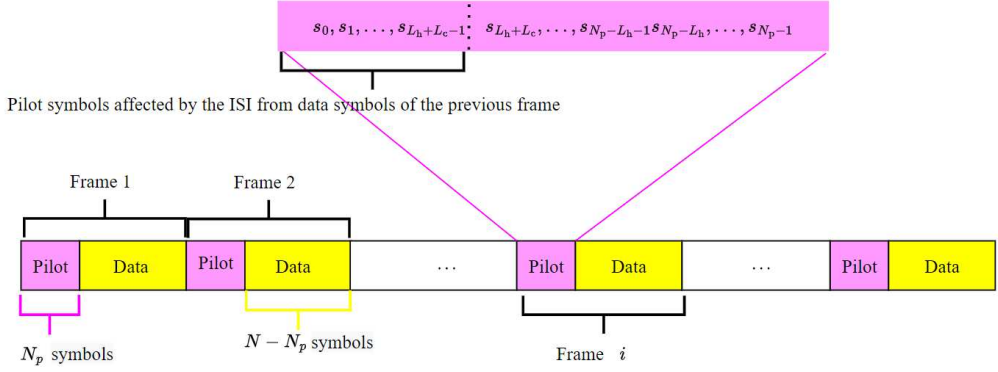}
\caption{The frame structure employed in the paper.}
\label{fig:3}
\end{figure*}

In theory, the FTN signaling ISI length is infinite. However, considering that the significant power of the raised cosine pulse is concentrated in and around its main lobe, we truncate the RRC pulse so that it has the length of $2L_\text{h}+1$. That said, the vector $\mathbf{h} = [h_{-L_\text{h}},h_{-L_\text{h}+1},\hdots,h_{L_\text{h}}]^\text{T}$ of length $2L_\text{h}+1$ represents the effective taps of the raised cosine pulse. The parameter $L_\text{h}$ is determined based on the roll-off factor of the RRC pulse and the FTN signaling packing ratio $\tau$.

\subsection{Adopted Frame Structure}
In this subsection, we discuss the proposed frame structure for FTN signaling taking the effects of IBI into consideration as shown in Fig. \ref{fig:3}. The sampled signal after the matched filter goes through a discrete-time whitening filter to decorrelate the colored noise samples $z_k$. In order to build the whitening filter, we use spectral factorization of the z-transform of $\mathbf{h}$,  i.e., $H(z) = V(z)V(1/z^\ast)$ \cite{prlja2008receivers}. This is equivalent to having a causal whitening filter $\textbf{v}$ of length $L_\text{h}$ such that $\textbf{v}\ast \underset{\bar{}}{\textbf{v}} = \textbf{h}$. 
Considering $\textbf{v}=[v_0, v_1, ..., v_{L_\text{h}-1}]^\text{T}$, we have $\underset{\bar{}}{\textbf{v}} = [v_{L_\text{h}-1},v_{L_\text{h}-2}, ..., v_0]^\text{T}$. 

The $k$th received sample in (\ref{equation7}) after passing through the whitening filter can be expressed as
\begin{IEEEeqnarray}{rcl}\label{equation7wh}
\tilde{r}_k &{}={}& \sum\limits_{l=0}^{L_\text{c}} c_{k,l} \sum\limits_{j=-\infty}^{k} s_{k-j} v_{j-l} +w_k,
\end{IEEEeqnarray}
where $w_k$ is the zero-mean white Gaussian noise with variance $\sigma_n^2$.

In order not to deteriorate the channel estimation quality, we design the pilot sequence to have two separated parts. As depicted in Fig. \ref{fig:3}, the $L_\text{h}+L_\text{c}$ symbols, ${s}_k, \, k=0, 1, \ldots, L_\text{h}+L_\text{c}-1$, at the beginning of the pilot sequence in the current frame constitute the first 
part, and they are affected by the ISI from the unknown data symbols of the previous frame as well as the pilot symbols of the current frame.  
Since, the first part of the pilot sequence is contaminated by ISI from the unknown data symbols, it will be discarded from further processing. Each symbol in the pilot sequence is also contaminated by the ISI from the previous and future pilot symbols. However, as the pilot symbols are known at the receiver, their ISI contribution can be easily mitigated, in contrast to the ISI from unknown data symbols. 
The symbols, ${s}_k, \, k=L_\text{h}+L_\text{c}, \ldots, N_\text{p}-1$, in the current frame constitute the second part which is the useful part of pilot sequence for channel estimation.

We define ${\tilde{p}_{k,j}}$ as the effective ISI resulting from both the frequency-selective channel and the FTN signaling as 
\begin{IEEEeqnarray}{rcl}\label{equation8wh}
\tilde{p}_{k,j}&{}={}&\sum\limits_{l=0}^{L_\text{c}} {c_{k,l} v_{j-l}}. \quad  j = 0,..., L_\text{h}+L_\text{c}-1.
\end{IEEEeqnarray}
This is as $v_{j-l}$ in (\ref{equation7wh}) is only nonzero where $(j-l)=0,\hdots, L_\text{h}-1$. In other words, $v_{j-l}$ is non-zero for 
$j = 0,..., L_\text{h}+L_\text{c}-1$. Thus, for $l = 0,\hdots, L_\text{c}$,  $h_{j-l}$ is zero for $j$ less than $0$ and $j$ greater than $L_\text{h}+L_\text{c}-1$. Thus, the convolution of $\textbf{v}$ of length $L_\text{h}$ and $(L_\text{c}+1)$-tap fading channel, results in an effective ISI of length $L_\text{eff} = L_\text{h}+L_\text{c}$. 
Considering (\ref{equation8wh}), we can rewrite (\ref{equation7wh}) for a given frame of received symbols, i.e., for $k=0,1,\hdots, N-1$, as
\begin{IEEEeqnarray}{rcl}\label{equation9wh}
\tilde{r}_k &{}={}& \sum\limits_{j=0}^{\min\left( 
L_\text{h}+L_\text{c}-1,k\right)} \tilde{p}_{k,j} s_{k-j}+w_k, \quad k=0, \hdots, N-1,   \IEEEeqnarraynumspace
\end{IEEEeqnarray}
where the summation limits result from the fact that the effective ISI $\tilde{p}_{k,j}$ in (\ref{equation8wh}) is zero for $j$ less than $0$ or $j$  greater than $L_\text{h}+L_\text{c}-1$.

\section{
Doubly-Selective Channel Estimation For FTN Signaling}
This section estimates the unknown doubly-selective channel impulse response with the aid of pilot sequences known to the receiver. Because of the induced ISI of FTN signaling, traditional Nyquist-based pilots cannot be directly applied to the FTN signaling case. We propose a novel channel estimation approach of FTN signaling to estimate the complex channel coefficients at the pilot locations within the frame. We also find the optimal pilot sequence that minimizes the MSE of the channel estimation. Finally, we employ a low-complexity linear interpolation to track the complex channel coefficients at the data symbols locations within the frame.

\subsection{Frequency-Selective Channel Estimation for FTN Signaling}

In this subsection, we process the received samples in (\ref{equation9wh}) corresponding to the transmitted pilot symbols to facilitate the
frequency-selective channel estimation. As discussed before and depicted in Fig. \ref{fig:3}, the $L_\text{h}+L_\text{c}$ symbols at the beginning of the pilot sequence, $s_k, k= 0,..., L_\text{h}+L_\text{c}-1$, suffer from the ISI from the unknown data symbols of the previous frame. 
Hence, in order to not deteriorate the channel estimation quality, these pilot symbols are discarded from further processing. Having said that, we process $r_k$, $k= L_\text{h}+L_\text{c},..., {N_\text{p}-1}$, which are the useful part of transmitted pilot sequence.

We design the duration of the pilot sequence to be less than the channel coherence time so that the channel coefficients can be regarded as constants over this duration. As a result, $c_{k,l}$ and $\tilde{p}_{k,j}$, $k= L_\text{h}+L_\text{c},..., {N_\text{p}-1}$ are not a function of $k$, and we can drop the time index $k$. We also design $N_\text{p}$ to be greater than the effective ISI length, i.e.,  $N_\text{p}>L_\text{eff}$, hence, the received samples in (\ref{equation9wh}) corresponding to the useful part of transmitted pilot symbols can be written as 
\begin{IEEEeqnarray}{rcl}\label{eqS1Wh}
\tilde{r}_k 
 &{}={}& \sum\limits_{j=0}^{L_\text{h}+L_\text{c}} \tilde{p}_{j} s_{k-j}  +\tilde{z}_k, \quad  k= L_\text{h}+L_\text{c},..., {N_\text{p}-1},
 \IEEEeqnarraynumspace
\end{IEEEeqnarray}
where
\begin{IEEEeqnarray}{rcl}\label{eqS0Wh}
\tilde{p}_{j}&{}={}&\sum\limits_{l=0}^{L_\text{c}} {c_{l} v_{j-l}}, \quad  j = 0,..., L_\text{h}+L_\text{c}.
\end{IEEEeqnarray} 
The received samples in (\ref{eqS1Wh}) can be rewritten as a vector form
\begin{IEEEeqnarray}{rcl}\label{equS2Wh}
 \mathbf{r}_\text{p} &{}={}& \mathbf{T} \mathbf{V} \mathbf{c} + \tilde{\textbf{z}}_\text{p},
\end{IEEEeqnarray}
where $\textbf{r}_\text{p}= [\tilde{r}_{L_\text{h}+L_\text{c}}, \tilde{r}_{L_\text{h}+L_\text{c}+1}, \ldots, \tilde{r}_{N_\text{p}-1}]^\text{T}$, $\tilde{\textbf{z}}_\text{p}=[\tilde{z}_{L_\text{h}+L_\text{c}},
\tilde{z}_{L_\text{h}+L_\text{c}+1} ... \tilde{z}_{N_\text{p}-1}]^\text{T}$ is the zero-mean white Gaussian noise with variance $\sigma_n^2$, and
$\mathbf{V} \in \mathbb{C}^{L_\text{eff} \times(L_\text{c}+1)}$ is the  circulant ISI matrix constructed from the elements of vector $\textbf{v}$ represented by $[v_{0}, v_{1},..., v_{L_\text{h}}]^\text{T}$,  $\textbf{c}=  [c_0, c_1, ..., c_{L_\text{c}}]^\text{T}$ is the unknown channel to be estimated, and $\mathbf{T} \in \mathbb{C}^{(N_\text{p}-L_\text{eff}+1)\times L_\text{eff}}$ is the matrix consists of pilot symbols and can be written as
\begin{IEEEeqnarray}{rcl}\label{eqS55noWh}
\mathbf{T} = 
\begin{bmatrix}
s_{L_\text{eff}-1}   & s_{L_\text{eff}-2} & \dots    & \dots      &s_{0}\\
s_{L_\text{eff}} & s_{L_\text{eff}-1}   & \dots     & \dots      &s_{1}\\
\vdots   & \vdots   & \ddots& \vdots  &\vdots\\
s_{N_\text{p}-1} & s_{N_\text{p}-2}   & \dots & \dots & s_{N_\text{p}-L_\text{eff}}\\
\end{bmatrix}.
\end{IEEEeqnarray}

To estimate the unknown channel $\mathbf{{c}}$ at useful part of the pilot sequence, i.e., at $k= L_\text{h}+L_\text{c},..., {N_\text{p}-1}$, the LSSE approach is employed by
minimizing the sum of squared errors as
\begin{IEEEeqnarray}{rcl}\label{equS5wh}
\mathbf{\hat{c}} &{}={}& \text{arg} \displaystyle \min_{\mathbf{c}}  (\mathbf{r}_\text{p}-\mathbf{T}   \mathbf{V}   \mathbf{c})^\text{H}(\mathbf{r}_\text{p}-\mathbf{T}   \mathbf{V} \mathbf{c})\nonumber\\
 &{}={}& (\mathbf{TV})^{-1} \mathbf{r}_\text{p}.
\end{IEEEeqnarray}
The matrix $\mathbf{T}$ in (\ref{equS5wh}) depends only on the pilot sequence which is known at the receiver side. As the vector $\mathbf{v}$ has constant taps for a given the roll-off factor of the RRC pulse and a given packing ratio of the FTN signaling, matrix $\mathbf{V}$ is also known at the receiver side. Thus, the matrix $(\mathbf{TV})^{-1} $ may be computed offline at the receiver. We select the pilot sequence in $\mathbf{T}$ that minimizes the MSE of the channel estimation in (\ref{equS5wh}). As $\tilde{\textbf{z}}_\text{p}$ is a zero-mean additive  Gaussian noise vector, then $\text{E}(\mathbf{\hat{c}}) = \mathbf{c}$, and $\mathbf{\hat{c}}$ is an the unbiased estimation of $\mathbf{c}$. Thus, the MSE of $\mathbf{\hat{c}}$ is equal to its variance and can be written as
\begin{IEEEeqnarray}{rcl}\label{equS6wh}
\text{MSE}
&{}={}& \text{tr} \left(\sigma_n^2 \mathbf{(TV)}^{-1} \left(\mathbf{(TV)}^{-1}\right)^\text{H}\right).
\end{IEEEeqnarray}
To design a pilot sequence suitable for channel estimation that takes the ISI caused by FTN signaling into account, the $\text{MSE}$ can be used as a criterion to compare the quality of different pilot sequences. To  this end, an exhaustive search can be done to find the pilot sequence, $\textbf{s}_\text{p} = [s_0, ..., s_{N_\text{p}-1}]^\text{T}$ that minimizes $\text{MSE}$.  This can be formally expressed as 
\begin{IEEEeqnarray}{rcl}\label{eqP2}
\textbf{s}_\text{p} &{}={}& \text{arg} \displaystyle \min_{\textbf{s}_\text{p} \in \mathcal{S}_\text{p}}  \text{MSE}.
\end{IEEEeqnarray}
where $\mathcal{S}_\text{p}$ indicates constellation set 
of pilot symbols.

\subsection{Tracking the Time-Selective Channel}

\begin{figure*}[t]
\centering
\includegraphics[width=0.75\textwidth]{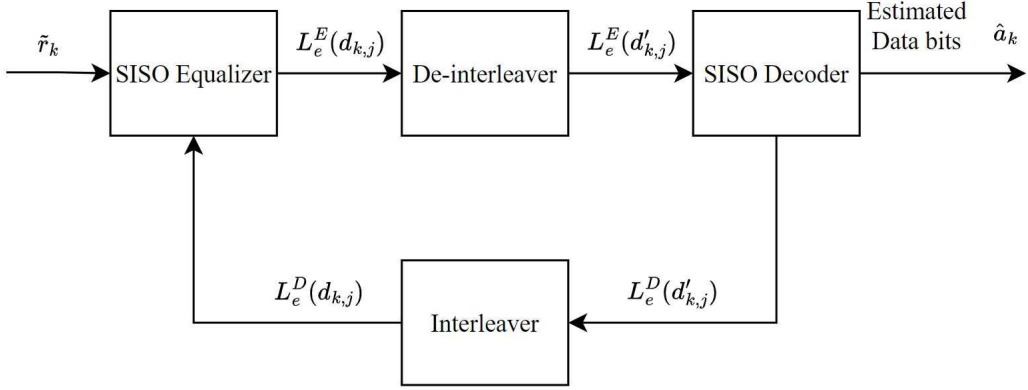}
\caption{Block diagram of a turbo equalizer.}
\label{fig:4}
\end{figure*}
For time-invariant channels, when the channel is estimated with the use of the pilot sequence in a given frame, the estimation can be considered to be constant during the entire data block in the same frame. That is to say, the  channel estimation can be done only once for each frame. However, for time-selective channels, the estimated channel at the pilot sequence location at the beginning of the frame can become obsoleted during the rest of the frame.  This will eventually increase the mean squared error that accumulates during the length of the frame.
The inaccuracy, resulting from assuming the same channel coefficients over the entire frame, can be addressed using an interpolation over the estimated coefficients at the pilot sequence located in the beginning of the current frame and the pilot sequence of the next frame.

The effects of three interpolation methods on the MSE of estimating the time-selective channel is investigated in Section \RomanNumeralCaps{5} in Fig. \ref{fig:77}. As can be seen from Fig. \ref{fig:77}, the MSE is almost the same for the different interpolation methods (linear, cubic, spline) employed to track the time-selective channel. Having said that, in our work, we adopt the linear interpolation due to its low-complexity. Let the channel estimation {at the pilot sequence at the beginning of the current frame, i.e., $i$th frame, and the pilot sequence of the next frame, i.e., $(i+1)$th frame} be $\mathbf{\hat{c}_i}$ and $\mathbf{\hat{c}_{i+1}}$, respectively. The channel estimation can be updated at each data symbol $k$, $k = 0,...,N_\text{d}-1$ within the $i$th frame, using the linear interpolation of $\mathbf{\hat{c}_i}$ and $\mathbf{\hat{c}_{i+1}}$ as
\begin{IEEEeqnarray}{rcl}\label{equInt}
\mathbf{\hat{c}}_{i,k+1} &{}={}& \mathbf{\hat{c}}_{i,k} + 
\frac{(\mathbf{\hat{c}_{i+1}}-\mathbf{\hat{c}_{i}})}{N_\text{d}}, \quad k=0,...,N_\text{d}-1, \IEEEeqnarraynumspace
\end{IEEEeqnarray}
where $\mathbf{\hat{c}}_{i,k}$ is the channel estimation for the $k$th data symbol in the $i$th frame and  $\mathbf{\hat{c}}_{i,0}$ is set to be $\mathbf{\hat{c}_i}$.

\begin{figure*}[t]
\centering
\includegraphics[width=0.75\textwidth]{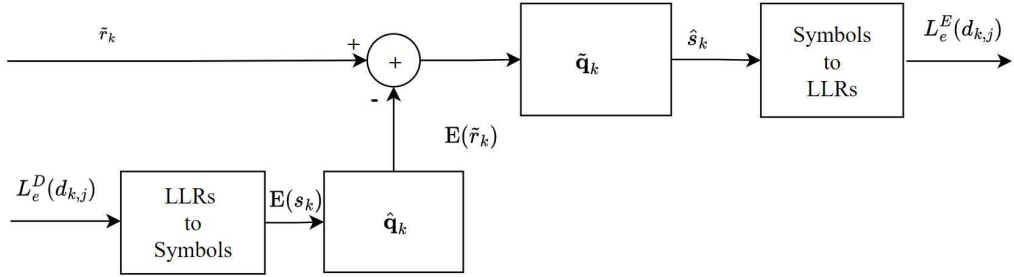}
\caption{Block diagram of a SISO equalizer.}
\label{fig:5}
\end{figure*}

\begin{figure*}[!t]
\normalsize
\setcounter{MYtempeqncnt}{\value{equation}}
\setcounter{equation}{20}
\begin{IEEEeqnarray}{rcl}\label{SISO31}
\tilde{\textbf{P}}_k &{}={}& 
\begin{bmatrix}
\tilde{p}_{k+L_1,0} & \tilde{p}_{k+L_1,1} & \dots  & \tilde{p}_{k+L_1,L_\text{h}+L_\text{c}} & 0 & \dots &0\\
0 & \tilde{p}_{k+L_1-1, 0} & \dots  &\tilde{p}_{k+L_1-1, L_\text{h}+L_\text{c}-1}  & \tilde{p}_{k+L_1-1,L_\text{h}+L_\text{c}} & \dots &0\\
\vdots& \vdots &\ddots &\vdots &\vdots &\vdots&\vdots\\
0  & 0   & \dots & \tilde{p}_{k-L_2,0}&\dots&\tilde{p}_{k-L_2,L_\text{h}+L_\text{c}-1}& \tilde{p}_{k-L_2,L_\text{h}+L_\text{c}}
\end{bmatrix}.\nonumber\\
\end{IEEEeqnarray}
\setcounter{equation}{\value{MYtempeqncnt}}
\hrulefill
\vspace*{4pt}
\end{figure*}

\section{FTN Signaling Detection Using Turbo Equalization}
This section summarizes the turbo equalizer based on a linear SISO MMSE equalizer to detect FTN signaling. In this section, we assume that the effective time-varying ISI introduced in (\ref{equation8wh}) is estimated and thus known at the receiver using the channel estimation method introduced in this paper.

The block diagram of the turbo equalizer is shown in Fig. \ref{fig:4}. The log-likelihood ratios (LLRs), which are the soft information of the coded bits, are exchanged between the SISO MMSE equalizer and the SISO decoder through a deinterleaver/interleaver pair in an iterative manner. After several equalization and decoding iterations, the quality of the soft information improves, and hard decisions from the SISO decoder can adequately estimate the transmitted data bits~\cite{otnes2002low}. 
The inputs to the SISO MMSE equalizer are the block of received samples, $\tilde{r}_k$, $k = N_\text{p},\hdots,N-1$,  and the extrinsic LLRs from the decoder denoted by $L_\text{e}^\text{D} (d_{k,j})$, where ``e" stands for extrinsic and ``D" stands for decoder. The SISO MMSE equalizer computes the soft information $L_\text{e}^\text{E}(d_{k,j})$ of the coded bits, where ``E" stands for the equalizer. Reversing the interleaving process in the transmitter can be done by deinterleaving $L_\text{e}^\text{E}(d_{k,j})$ into $L_\text{e}^\text{E}({d'_{k,j}})$, which is employed as a priori information to be fed into the decoder~\cite{otnes2003evaluation}.
The SISO decoder calculates a posteriori LLRs, $L_D({d'_{k,j}})$, which can be used to compute the extrinsic LLRs as ${L_\text{e}}^\text{D}({d'_{k,j}})=L_\text{D} ({d'_{k,j}})-L_\text{e}^\text{E}({d'_{k,j}})$. The decoder also outputs $\hat{a}$ as an estimation of the transmitted data bits. The extrinsic LLRs of the decoder ${L_\text{e}}^\text{D}({d'_{k,j}})$ then interleaved into $L_\text{e}^\text{D}(d_{k,j})$ and fed back to the SISO equalizer in the next iteration~\cite{otnes2002low}.

Please note that the a priori LLRs, $L_\text{e}^\text{D}(d_{k,j})$, input to the MMSE equalizer are zero in the first equalization and decoding step, namely the zero-th iteration. The pilot symbols, which are known at the receiver, are considered to be a perfect a priori information in all iterations, including the zero-th iteration.

\subsection{The SISO Equalizer}
This subsection reviews the design of the optimal coefficient of a filter-based SISO MMSE equalizer, illustrated in Fig. \ref{fig:5}. The SISO MMSE equalizer is a time-varying filter $\tilde{\textbf{q}}_k = [\tilde{q}_{L_1,k}, \tilde{q}_{L_1-1,k}, \hdots ,\tilde{q}_{-L_2,k}]^\text{T}$ having the length $L_\text{eq}$ such that $L_\text{eq}=L_1+L_2+1$. To estimate transmitted data symbols of the current frame from the received data symbols after whitening filter $\tilde{r}_k$, $k= N_\text{p},..., {N-1}$, the received samples after whitening filter in (\ref{equation9wh}) corresponding to the data symbols can be rewritten as 
\begin{IEEEeqnarray}{rcl}\label{SISO1wh}
\tilde{r}_k &{}={}& \sum\limits_{j=0}^{L_\text{h}+L_\text{c}-1} \tilde{p}_{k,j} s_{k-j}+w_k,  \quad k = N_\text{p},\hdots,N-1,\IEEEeqnarraynumspace
\end{IEEEeqnarray}
where
\begin{IEEEeqnarray}{rcl}\label{SISO111wh}
\tilde{p}_{k,j} &{}={}& \sum\limits_{l=0}^{L_\text{c}} {c_{k,l} h_{j-l}}.  \quad j = 0,\ldots, L_\text{h}+L_\text{c}-1,
\end{IEEEeqnarray}
Given the length of the equalizer filter, an estimation of transmitted symbol $\hat{s}_k$ can be calculated from the $L_\text{eq}$ received samples $\tilde{r}_k$ and a priori information on the $L_\text{eq}+L_\text{eff}$ transmitted symbols $s_k$. Defining $\tilde{\textbf{r}}_k=[\tilde{r}_{k+L_1},\tilde{r}_{k+L_1-1},\ldots,\tilde{r}_{k-L_2}]^T$, as a vector of received symbols at the equalizer introduced in (\ref{SISO1wh}), $\tilde{\textbf{r}}_k$ may be written in matrix format as
\begin{IEEEeqnarray}{rcl}\label{SISO22}
\tilde{\textbf{r}}_k &{}={}& \tilde{\textbf{P}}_k \textbf{s}_k +\textbf{w}_k,  \quad  k=N_\text{p}, \hdots, N-1,      
\end{IEEEeqnarray}
where $\textbf{s}_k=[s_{k+L_1+L_\text{h}},s_{k+L_1+L_\text{h}-1},\ldots,s_{k-L_2-L_\text{h}-L_\text{c}}]^\text{T}$ is the vector of transmitted symbols and $\tilde{\textbf{P}}_k \in \mathbb{C}^{L_\text{eq} \times (L_\text{eq} + L_{\text{eff}} )}$ is the effective ISI matrix defined in (\ref{SISO31}) at the top of next page.
\addtocounter{equation}{1}
The $\textbf{w}_k=[w_{k+L_1},w_{k+L_1-1},\ldots,w_{k-L_2}]^\text{T}$ is a vector of zero-mean white noise samples with covariance $\text{Cov}(\textbf{w}_k,\textbf{w}_k) = \sigma_n^2 \textbf{I}$. The MMSE criteria in which the mean square error $\text{E}({|s_k-\hat{s_k}|}^2)$ is minimized, is employed to calculate the equalizer coefficients ${\tilde{\textbf{q}}_k}$. 
As introduced in \cite{tuchler2002minimum}, the optimal linear MMSE estimate $\hat{s}_k$ of ${s}_k$ is expressed as 
\begin{IEEEeqnarray}{rcl}\label{SISO13wh}
\hat{s}_k &{}={}&  {\tilde{\textbf{q}}_k}^{\text{H}} (\tilde{\textbf{P}}_k-\textbf{r}_k \text{E}(\textbf{s}_k)+\tilde{\textbf{p}}_k \text{E}({s}_k)),
\end{IEEEeqnarray} 
where $\tilde{\textbf{p}}_k = \tilde{\textbf{P}}_k \textbf{u}_k$
Thus, the optimal filter coefficient, $\tilde{\textbf{q}}_k$ can be found as \cite{tuchler2002minimum}
\begin{IEEEeqnarray}{rcl}\label{SISO14wh}
\tilde{\textbf{q}}_k &{}={}& {\left(\tilde{\textbf{P}}_k \Sigma_k {\tilde{\textbf{P}}_k}^{\text{H}}+\sigma_n^2 \textbf{I} +(1-\sigma_k)\tilde{\textbf{p}}_k {\tilde{\textbf{p}}_k}^{\text{H}}\right)}^{-1} \tilde{\textbf{p}}_k.
\end{IEEEeqnarray}

where ${\Sigma}_k=\text{Cov}(\textbf{s}_k,\textbf {s}_k) =\text{Diag}([\sigma_{k+L_1+L_\text{h}},\sigma_{k+L_1+L_\text{h}-1},\ldots,\sigma_{k-L_2-L_\text{h}-L_\text{c}}])$, $\textbf{u}_k=[\textbf{0}_{L_1+L_\text{h}},1,\textbf{0}_{L_2+L_\text{h}+L_\text{c}}]^\text{T}$ and variance $\sigma_k= \text{E}({s}_k {s}^*_k)-\text{E}({s}_k)\text{E}({s}^*_k)$.
A priori mean $\text{E}({s}_k)$ and variance $\sigma_k$ can be computed from the ${L_\text{e}}^\text{D}(c_{k,j})$ as discussed in \cite{tuchler2002minimum}. One can calculate the LLRs $L_\text{e}^\text{E} (c_{k,j})$ from $\hat{s}_k$ \cite{tuchler2002minimum} and \cite{otnes2003evaluation}.
It should be noted that calculating the optimal filter coefficient, $\tilde{\textbf{q}}_k$, has a computational complexity per time step proportional to $N^3$. Fortunately, a time-recursive algorithm is introduced in~\cite{otnes2002low} with a computational complexity per time step proportional to $N^2$.

\section{Simulation Results}

This section provides the numerical results of the MSE and BER performance to evaluate the proposed channel estimation and FTN signaling detection schemes. 

\subsection{Simulation Setup}
{On the transmitter side, we set $T$ to $0.41667$ ms and use a rate $1/2$ convolutional code with a constraint length $7$ and the two generator polynomials 0x5b and 0x79}. A code rate of 3/4 from the rate 1/2 code can be achieved through puncturing the output bits of the encoder with a puncturing mask of $[1 1 1 0 0 1]$, where a “1” and a “0” indicate that the bit is transmitted and discarded, respectively. A posteriori probability (APP) decoding algorithm is employed at the receiver to decode the convolutional code. 
We consider binary phase shift keying (BPSK) modulation for both pilot and data symbols and a RRC pulse with a roll-off factor of $\beta = 0.35$ unless otherwise mentioned. 

{We consider three channel models: 1) ITU-R Poor channel, a doubly-selective Rayleigh fading channel with fading rate of $0.0004$ and delay spread of 2.1 ms, and it will be denoted as \textit{channel model 1}. Hence, we set $N_\text{p}$ and $N_\text{d}$ to 32 and 256 symbols, respectively. 
2) a frequency-selective Rayleigh fading channel described with similar delay spread of channel model 1 whose impulse response remains constant over two successive transmission frames, and it will be denoted as \textit{channel model 2}.
3) a frequency-selective Rayleigh fading channel having a delay spread of $L_\text{c} = 8$ whose impulse response remains constant over two successive transmission frames, and it will be denoted as \textit{channel model 3}. Please note that \textit{channel model 3} is mainly considered to have a fair comparison with \cite{ishihara2017iterative}. High frequency (HF) communication is an effective method of communicating over very long distances in rural and remote areas \cite{wang2018hf}. Having said that we employ the the HF model with poor channel conditions as recommended by ITU-R. The ITU-R Poor channel consists of two independent but equal average power Rayleigh fading paths with a fixed 2 ms delay between paths and with a frequency Doppler of 1 Hz. We consider the 2.1 ms delay spread for the channel to have exactly 5 symbol intervals in the Nyquist case having a 2400 symbol rate which is equivalent to a Watterson model with 6 taps \cite{otnes2003evaluation}. This is equivalent to the fading rate of 0.0004. The (\textit{channel model 2}) is the time-invariant version of the \textit{channel model 1}.}

We adopt the following notation for the considered cases through the simulation results section:
\begin{itemize}
 \item \textit{F-F} to denote the case of employing the proposed FTN pilot sequence in the FTN signaling transmission.
 \item \textit{N-F} for the case of employing the conventional Nyquist pilot sequence designed in \cite{crozier1991least} in the FTN signaling transmission.
 \item \textit{N-N} for the case of employing the conventional Nyquist pilot sequence introduced in \cite{crozier1991least} in the Nyquist signaling transmission. This case is used as a benchmark.
\item {Numerical\textit{-F} to denote the case of employing the FTN pilot sequence found by the interior-point algorithm minimizing the  MSE in (\ref{equS6wh}) for the FTN signaling transmission.}

\end{itemize}

 \begin{figure}[t!]
\centering
\includegraphics[width=8cm,height=6cm]{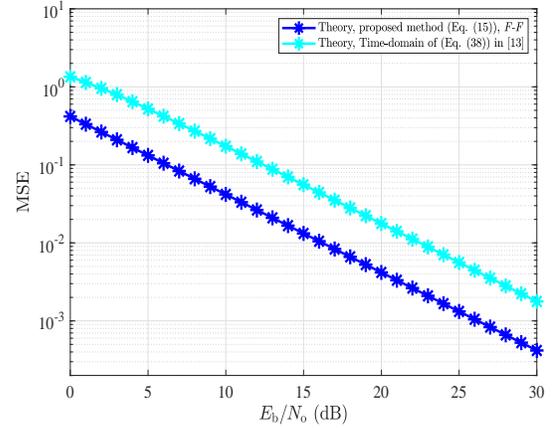}
\caption{The MSE of the proposed FTN signaling channel estimation employing the designed optimal pilot sequence versus the method in \cite{ishihara2017iterative} over \textit{channel model 3} for $\tau = 0.8$.}
\label{fig:6_0}
\end{figure}

 \begin{figure}[t!]
\centering
\includegraphics[width=8cm,height=6cm]{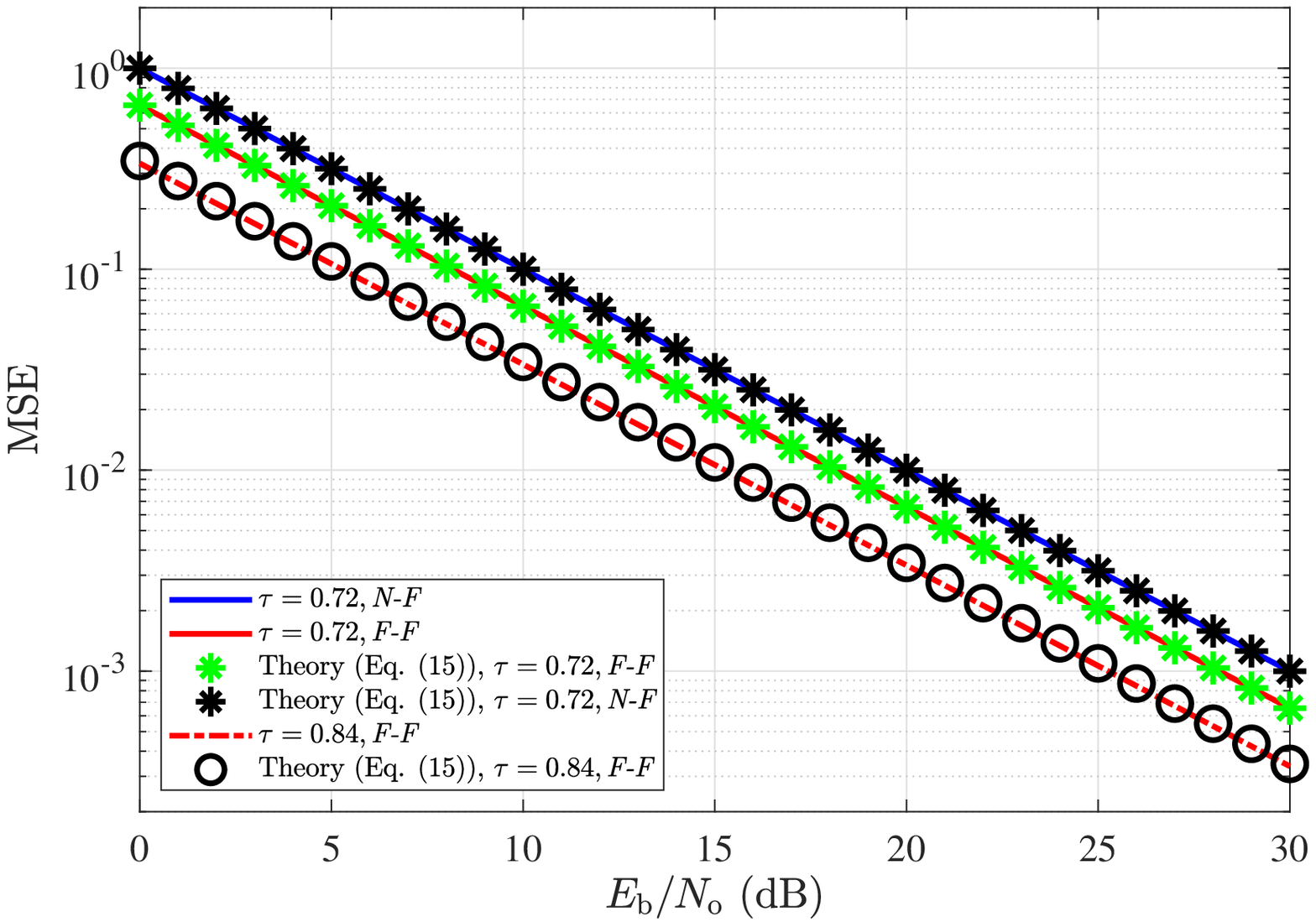}
\caption{The MSE of the proposed channel estimation for FTN signaling with $\tau = 0.72$ and $\tau = 0.84$ over \textit{channel model 2}.}
\label{fig:6}
\end{figure}

 \begin{figure}[t!]
\centering
\includegraphics[width=8cm,height=6cm]{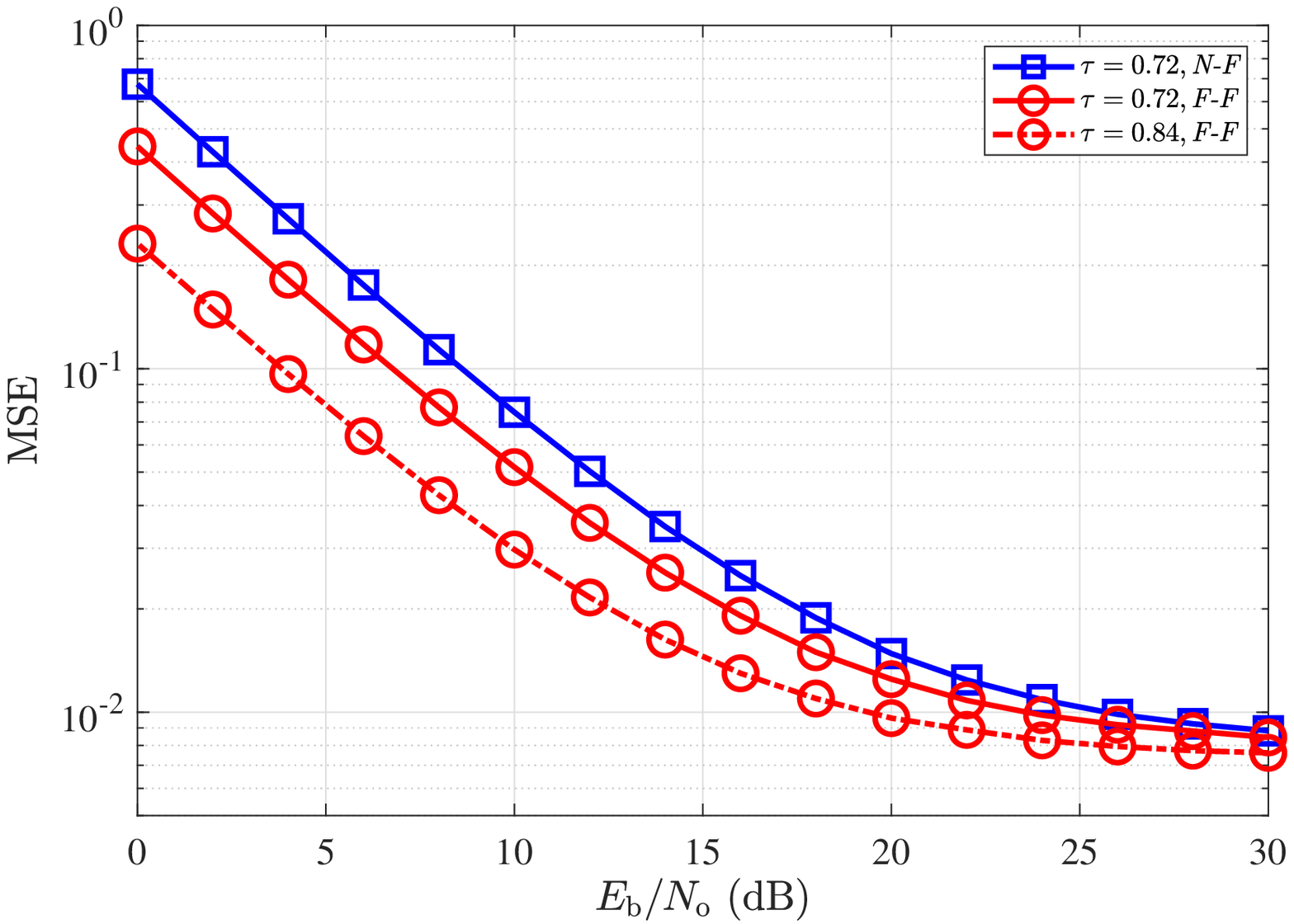}
\caption{The MSE of the proposed channel estimation for FTN signaling with $\tau =0.72$ and $\tau = 0.84$ over \textit{channel model 1}.}
\label{fig:7}
\end{figure}

 \begin{figure}[t!]
\centering
\includegraphics[width=8cm,height=6cm]{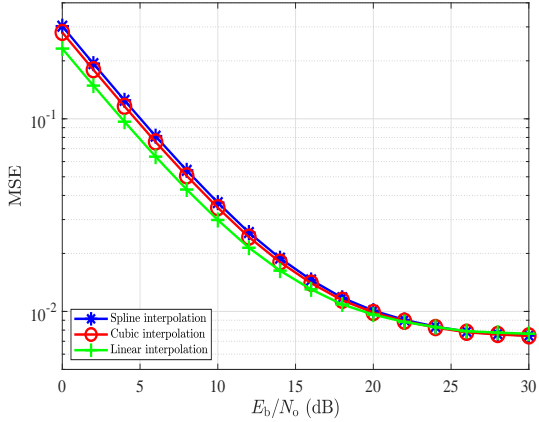}
\caption{The MSE performance of employing different interpolation methods for the proposed channel estimation for FTN signaling with $\tau = 0.84$ over \textit{channel model 1}.}
\label{fig:77}
\end{figure}

\begin{figure}[t!]
\centering
\includegraphics[width=8cm,height=6cm]{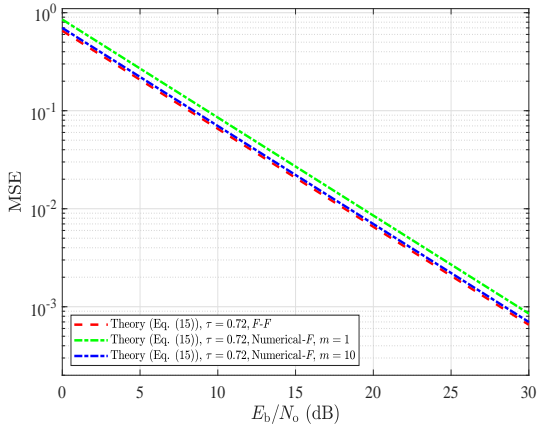}
\caption{The MSE performance of FTN signaling channel estimation employing the designed optimal pilot sequence found through exhaustive search versus the case using numerical optimization to find the pilot with $\tau =0.72$ over \textit{channel model 2}.}
\label{fig:30}
\end{figure}

In our simulations, in every simulation run, a superframe consisting of 144 frames is sent over the channel. The MSE in a given simulation run is calculated as
$\sum\nolimits_{l=0}^{L_{c}} ({\hat{\textbf{c}}_l}-\textbf{c}_l)^\text{H}({\hat{\textbf{c}}_l}-\textbf{c}_l)$, and the final simulated MSE is averaged over the total number of simulation runs. The MSE simulation results is also compared to the theoretical MSE in (\ref{equS6wh}). 

\subsection{Evaluation of the MSE of the proposed channel estimation}
Fig. \ref{fig:6_0} depicts the MSE of the proposed channel estimation employing the designed optimal pilot sequence
for FTN signaling having $\tau = 0.8$ over \textit{channel model 3}. {To be consistent with the \cite{ishihara2017iterative}, $\beta=0.5$ is considered for Fig. \ref{fig:6_0}. In all the simulation results except the ones that we compare our work with \cite{ishihara2017iterative} (Fig. \ref{fig:6_0} and Fig. \ref{fig:30}), $\beta$ is set to $0.35$ which is a practical value in communication systems.} To have a fair comparison with \cite{ishihara2017iterative}, we fix the SE of both transmission to 0.7407 bits/s/Hz. This is achieved as follows. The SE in our work is defined as $\text{SE} = \frac{N_\text{d}}{(N_\text{d}+N_\text{p})} \cdot\frac{{\log_2}M}{\tau (1+\beta)} = \frac{256}{(256+32)} \cdot \frac{{\log_2}2}{0.8 (1+0.5)} = 0.7407$ bits/s/Hz. To achieve the same SE, we set the pilot sequence length in \cite{ishihara2017iterative} to $8$. This is because a two pilot sequence length is added at the beginning and at the end of each data block to construct a frame in \cite{ishihara2017iterative}. {That is to say taking the pilot length to be 8 in \cite{ishihara2017iterative} results in having 32 known symbols (overhead of 32) in each frame, the same as the overhead in a given frame in our paper.} 
{The MSE of \cite{ishihara2017iterative} is expressed in the frequency domain (\cite{ishihara2017iterative}, Eq. (38)), and hence, to have a fair comparison of the MSE in our work, we convert the MSE in \cite{ishihara2017iterative} to the time-domain. It should be noted that the time-domain MSE is also equivalent to calculating $\sum\nolimits_{l=0}^{L_{c}} ({\hat{\textbf{c}}_l}-\textbf{c}_l)^\text{H}({\hat{\textbf{c}}_l}-\textbf{c}_l)$.} As depicted in Fig. \ref{fig:6_0}, the proposed FTN signaling channel estimation employing the designed optimal pilot sequence in this paper results in a significantly better (about 6 dB) MSE when compared to the MSE resulting from the channel estimation method employing designed pilot sequence in \cite{ishihara2017iterative}.
{It is worth mentioning that the emphasis in this paper is on the pilot design and the channel estimation rather than iterative channel estimation and FTN signaling detection.}

Fig. \ref{fig:6} depicts the MSE of the proposed channel estimation for FTN signaling having $\tau = 0.72$ and $\tau = 0.84$ over \textit{channel model 2}. Particularly, the \textit{F-F}, \textit{N-F} cases for $\tau = 0.72$ and \textit{F-F} case for $\tau = 0.84$ are considered. 
Fig. \ref{fig:6} depicts the simulation results for MSE, $\sum\nolimits_{l=0}^{L_{c}} ({\hat{\textbf{c}}_l}-\textbf{c}_l)^\text{H}({\hat{\textbf{c}}_l}-\textbf{c}_l)$, and the theoretical MSE in (\ref{equS6wh}). 
As illustrated in Fig. \ref{fig:6}, the MSE resulting from the simulation for all the employed pilot sequences matches the theoretical results of the MSE in (\ref{equS6wh}). 
Fig. \ref{fig:6} compares \textit{F-F} for the FTN signaling system having $\tau = 0.72$ with $\tau = 0.84$. As expected, the MSE for $\tau = 0.84$ is lower compared with $\tau = 0.72$. As it is shown in Fig. \ref{fig:6}, by increasing the $E_\text{b}/N_0$, the MSE decreases.
As depicted in Fig. \ref{fig:6} for $\tau = 0.72$, the MSE for \textit{F-F} case is about 2 dB better than \textit{N-F} case. That is to say, the optimal pilot sequence designed in this paper achieves a significantly better MSE compared to the Nyquist pilot sequence used in the FTN transmission signaling. 

The MSE performance of the proposed channel estimation for FTN signaling having $\tau = 0.72$ and $\tau = 0.84$ over \textit{channel model 1} is depicted in Fig. \ref{fig:7}. Particularly, the \textit{F-F} and \textit{N-F} cases for $\tau = 0.72$ and \textit{F-F} case for $\tau = 0.84$ are considered. 
As depicted in Fig. \ref{fig:7} for the $\tau = 0.72$, the MSE for \textit{F-F} cases are about 2 dB (for MSE of $9\times10^{-3}$) better than \textit{N-F} case. That is to say, the optimal pilot sequence designed in this paper achieves a significantly better MSE compared to the Nyquist pilot sequence used in the FTN transmission signaling. Fig. \ref{fig:7} compares \textit{F-F} for the FTN signaling system having $\tau = 0.72$ with $\tau = 0.84$ over \textit{channel model 1}. As expected, the simulation shows that the MSE for $\tau = 0.84$ is lower compared with $\tau = 0.72$ case. 
As it is shown in Fig. \ref{fig:6} for FTN signaling over \textit{channel model 2}, by increasing the $E_\text{b}/N_0$, the MSE decreases. However, as \textit{channel model 1} a doubly-selective channel, is employed in Fig. \ref{fig:7}, the MSE is saturated in high $E_\text{b}/N_0$. This is because of the interpolation error to track the time-selective channel.

Fig. \ref{fig:77} depicts the MSE of the proposed channel estimation for FTN signaling with $\tau = 0.84$ over \textit{channel model 1} in case of employing different interpolation methods. As it is shown in Fig. \ref{fig:77}, the MSE does not change significantly for the different interpolation methods employed to track the time-selective channel.

{In the above MSE performance simulations, we employed an exhaustive search to find the optimal pilot minimizing the MSE in (\ref{equS6wh}). This is mainly as pilot design can be carried out offline and only once for a given channel. Additionally, the pilot design optimization problem is non-convex, and low-complexity solutions are not necessarily guaranteed to reach the global optimal solution which may harm the channel estimation quality and degrate the error rate. To strike a balance between MSE performance and computational complexity, we relax the integer constraint of the problem in (\ref{equS6wh}) and solve the relaxed optimization problem numerically using the interior-point algorithm \cite{wright1997primal} to obtain a sub-optimal solution. This sub-optimal solution is then rounded to find the final pilot sequence. Please note that due rounding the continuous sub-optimal solution, optimality is no longer guaranteed. Additionally, given the non-convex nature of the optimization problem, the interior-point algorithm may stuck at a local optimal solution. To increase our chances of getting good quality local sub-optimal solution, we initialize the interior-point algorithm with value of $m$ uniformly distributed random pilot sequences and calculated the MSE and then choose the sub-optimal pilot sequence with the lowest MSE value. The simulation is done for $m = 1$ and $10$. As depicted in the Fig. \ref{fig:30}, by increasing the $m$, the MSE approaches the optimal MSE calculated using the exhaustive search. That is to say the interior-point algorithm, which has polynomial time complexity \cite{wright1997primal}, seems a promising approach to find sub-optimal pilots.}

\subsection{BER Performance Evaluation}
\begin{figure}[t!]
\centering
\includegraphics[width=8cm,height=6cm]{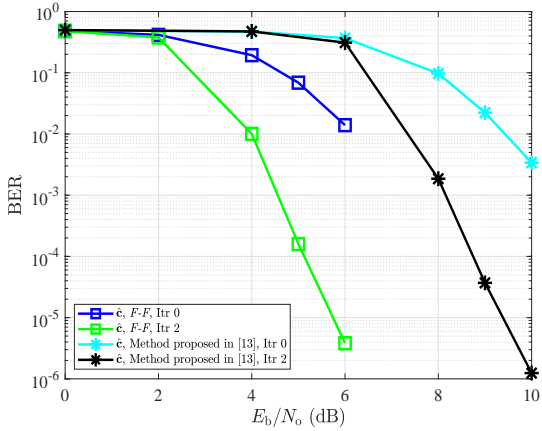}
\caption{The BER performance of FTN signaling channel estimation employing the designed optimal pilot sequence versus the method in \cite{ishihara2017iterative} with $\tau =0.8$ over \textit{channel model 3}.}
\label{fig:20}
\end{figure}

\begin{figure}[t!]
\centering
\includegraphics[width=8cm,height=6cm]{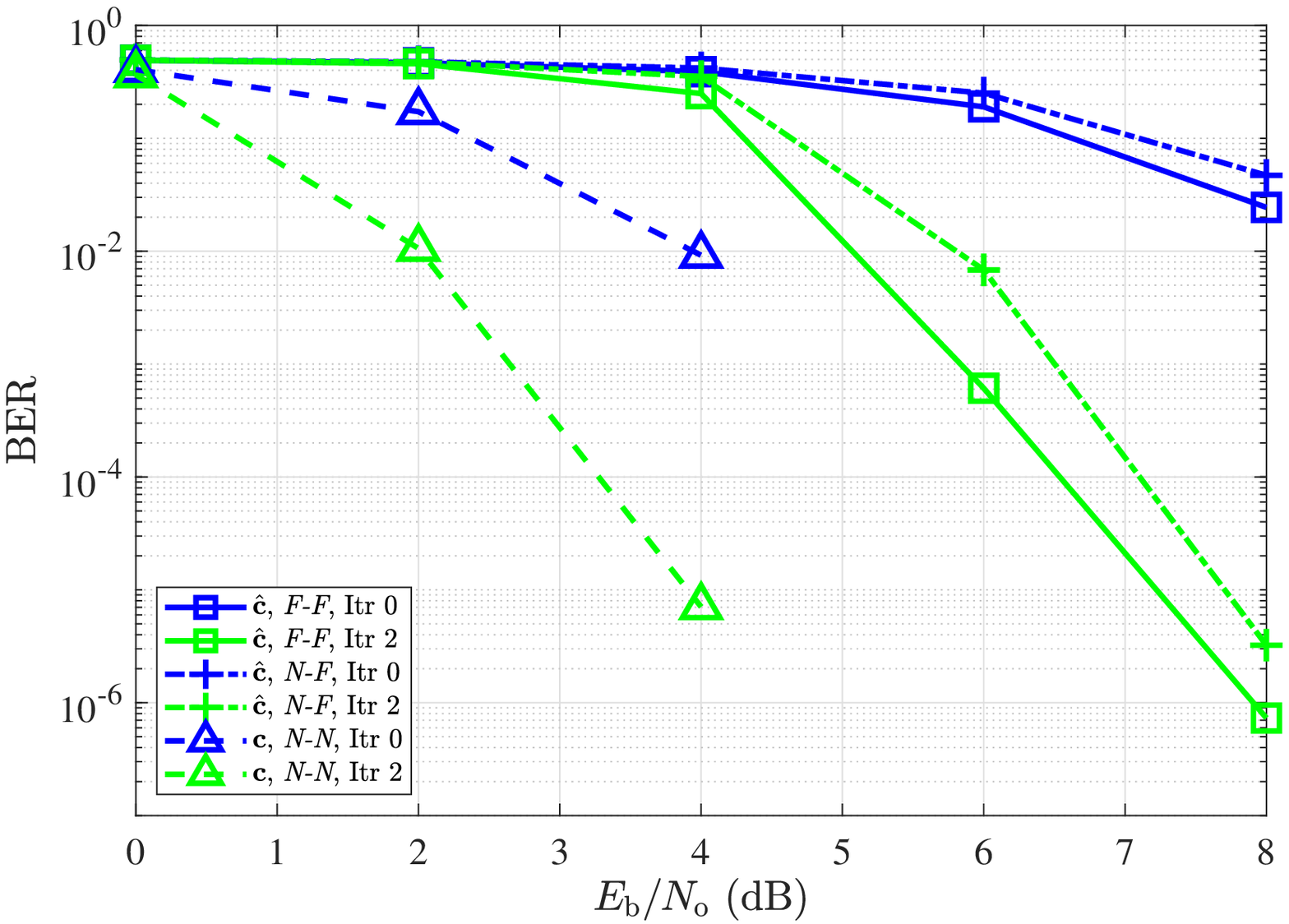}
\caption{The BER performance of FTN signaling with $\tau =0.72$ over \textit{channel model 2}.}
\label{fig:8}
\end{figure}

\begin{figure}[t!]
\centering
\includegraphics[width=8cm,height=6cm]{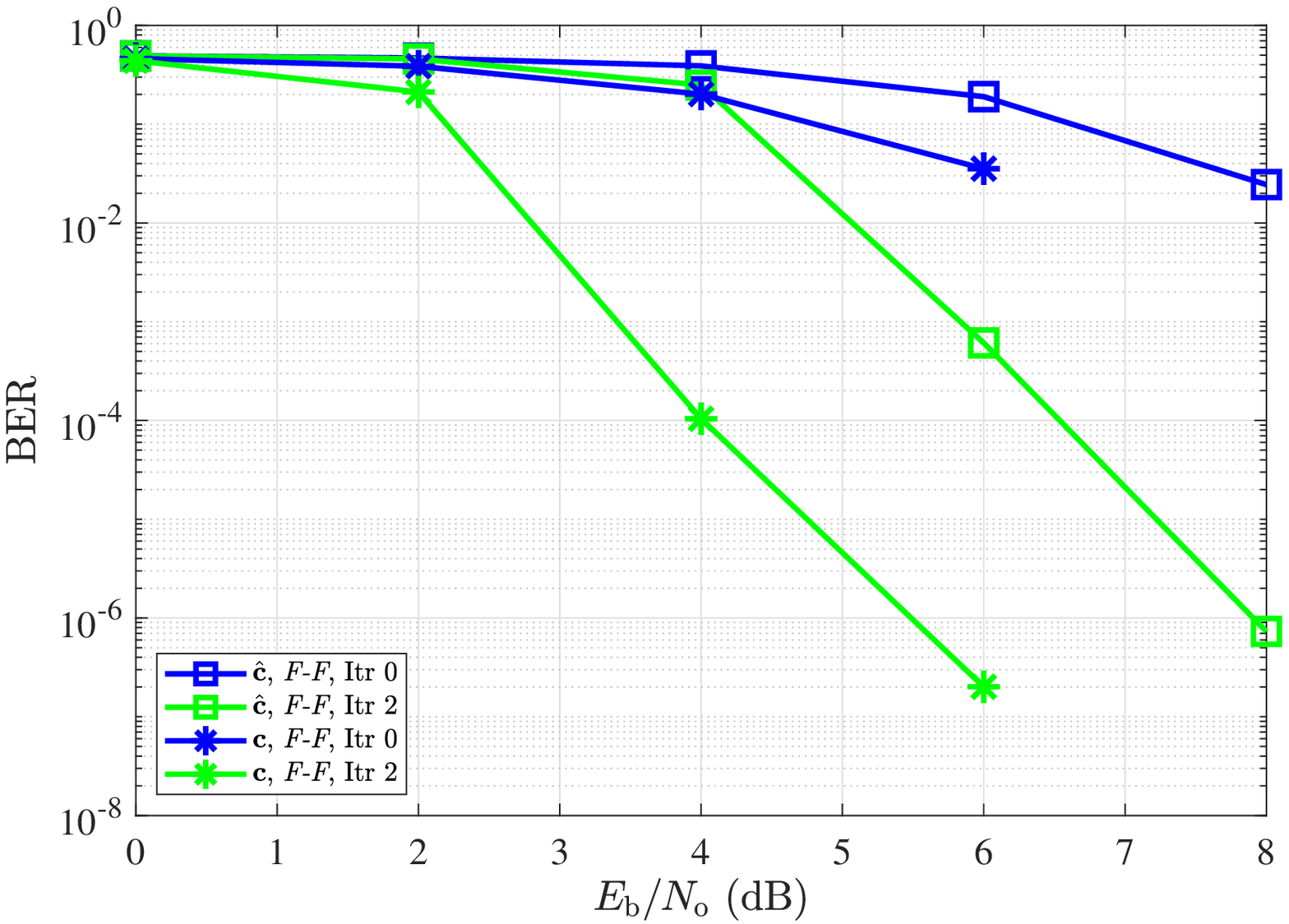}
\caption{The BER performance of FTN signaling with $\tau =0.72$ over \textit{channel model 2}.}
\label{fig:9}
\end{figure}

\begin{figure}[t!]
\centering
\includegraphics[width=8cm,height=6cm]{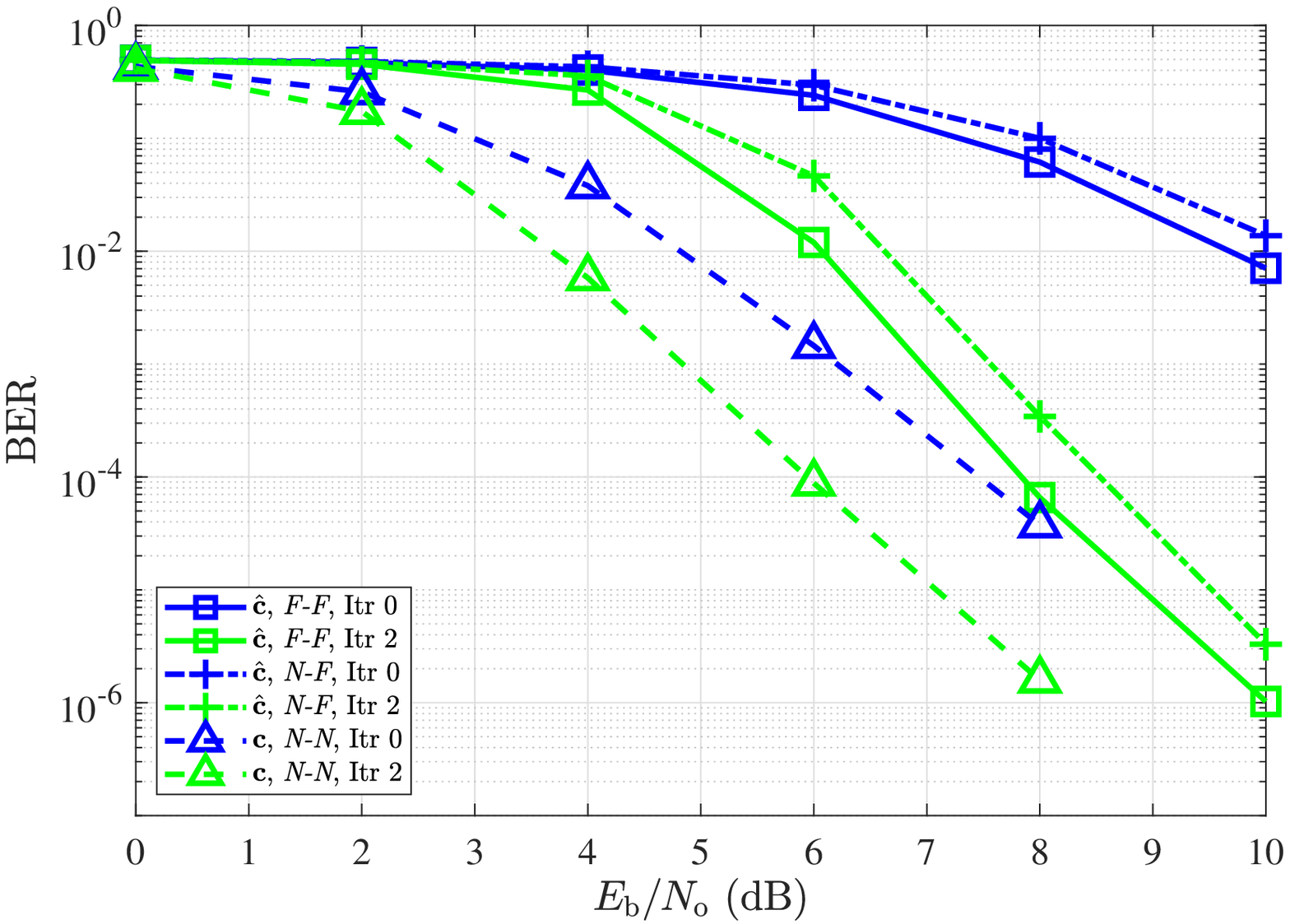}
\caption{The BER performance of FTN signaling with $\tau =0.72$ over \textit{channel model 1}.}
\label{fig:10}
\end{figure}

\begin{figure}[t!]
\centering
\includegraphics[width=8cm,height=6cm]{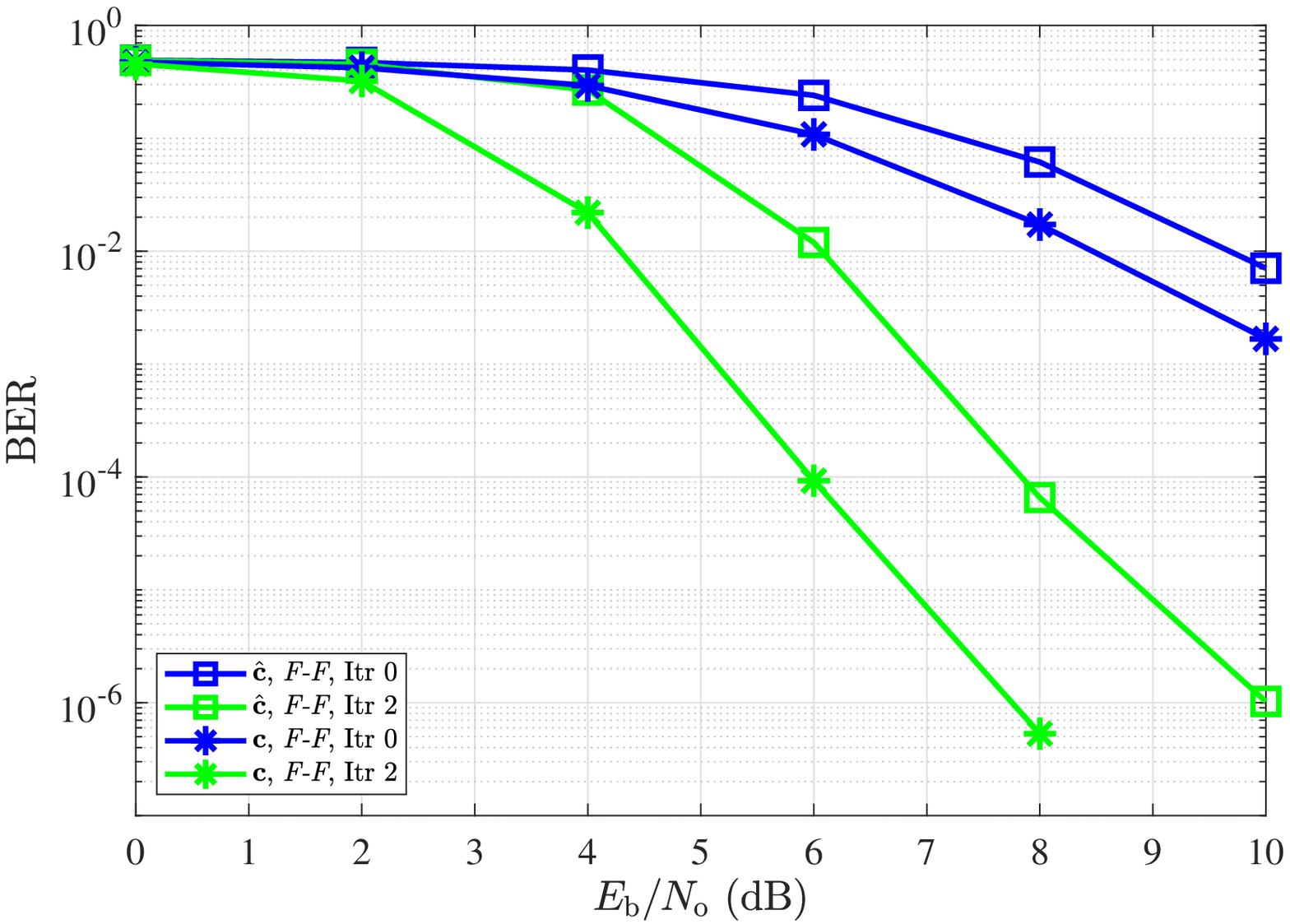}
\caption{The BER performance of FTN signaling with $\tau =0.72$ over \textit{channel model 1}}.
\label{fig:11}
\end{figure}

{Fig. \ref{fig:20} depicts the BER performance of the proposed channel estimation employing the designed optimal pilot sequence of the FTN signaling detection using \textit{F-F} having $\tau = 0.8$ over \textit{channel model 3}. To be consistent with \cite{ishihara2017iterative}, $\beta=0.5$ is considered. To guarantee  a fair comparison with \cite{ishihara2017iterative}, we fix the SE of both transmission to 0.7407 bits/s/Hz. The channel estimation method in \cite{ishihara2017iterative} is implemented and the BER output of the SISO decoder is measured to have a fair comparision. As depicted in Fig. \ref{fig:20}, the proposed FTN signaling channel estimation employing the designed optimal pilot sequence in this paper shows better BER compared to the channel estimation method employing the designed pilot sequence in \cite{ishihara2017iterative}. More specifically, after two iterations, the BER achieved employing the proposed channel estimation method using the designed optimal pilot sequence shows about 3.5 dB improvement compared to the case of \cite{ishihara2017iterative}.}

Fig. \ref{fig:8} compares the BER performance of the proposed channel estimation of the FTN signaling detection using \textit{F-F} and \textit{N-F} with $\tau = 0.72$ over \textit{channel model 2}. As a benchmark, the Nyquist case, \textit{N-N}, with perfectly known channel at the receiver is plotted.
As depicted in Fig. \ref{fig:8}, the BER for \textit{F-F} case is lower than \textit{N-F}. That is to say, the designed optimal pilot sequence in this paper achieves a better BER compared to the Nyquist pilot sequence used in the FTN transmission signaling. More specifically, after two iterations, the BER achieved using the designed optimal pilot sequence shows about 0.5 dB improvement compared to the case employing the Nyquist pilot sequence. As illustrated in Fig. (\ref{fig:8}), after two iterations, the BER achieved using the proposed channel estimation and designed optimal pilot sequence for FTN signaling system \textit{F-F} shows about 3 dB difference with the Nyquist scenario \textit{N-N} where channel is completely known at the receiver.


Fig. \ref{fig:9} compares the BER performance of the proposed channel estimation for \textit{F-F} and the case with perfectly known channel at the receiver in a FTN signaling system with $\tau = 0.72$ over \textit{channel model 2}. As shown in Fig. \ref{fig:9}, after only one iteration, the BER achieved by the proposed channel estimation employing the designed optimal pilot sequence in this paper shows about 2 dB difference with the case where channel is perfectly known at the receiver.


Fig. \ref{fig:10} compares the BER performance of \textit{F-F} and \textit{N-F} scenarios in a FTN signaling system with $\tau = 0.72$ over \textit{channel model 1}. As a benchmark, the \textit{N-N} case with perfectly known channel at the receiver is also illustrated. It is shown in Fig. \ref{fig:8} that \textit{F-F} achieves better BER compared to the \textit{N-F}. That is to say, the designed optimal pilot sequence achieves better BER comparing to the Nyquist pilot sequence.
As depicted in Fig. \ref{fig:10}, after two iterations, the BER achieved of the proposed channel estimation employing designed pilot sequence for FTN signaling system \textit{F-F} shows about 2 dB difference with the Nyquist scenario \textit{N-N} where channel is completely known at the receiver.

Fig. \ref{fig:11} compares the BER performance of the proposed channel estimation for \textit{F-F} and the case with perfectly known channel at the receiver in FTN signaling system with $\tau = 0.72$ over \textit{channel model 1}. As depicted in Fig. \ref{fig:11}, for the FTN signaling system, after two iterations, the BER achieved employing the proposed channel estimation and pilot design method in the paper shows less than 2 dB difference with the case where channel is perfectly known at the receiver.

\section{Conclusion}

In this paper, we proposed a channel estimation and data detection for FTN signaling over doubly-selective channels without imposing additional SE loss. To improve the channel estimation quality, we adopted a frame structure considering the ISI introduced by both FTN signaling and the frequency-selective channel. We proposed a LSSE-based channel estimation approach for FTN signaling to estimate the complex channel coefficients at the pilot locations within the frame. To this end, the optimal FTN signaling pilot sequence that minimized the MSE of the channel estimation is found. A low-complexity linear interpolation was employed to track the time-selective complex channel coefficients at the data symbols locations within the frame. A turbo equalization technique based on a linear SISO MMSE equalizer was employed to detect the data symbols of FTN signaling.
Simulation results showed that employing the designed optimal pilot sequence resulted in a better MSE of the channel estimation (up to 2 dB for both doubly-selective and frequency-selective channel models) compared to the Nyquist pilot sequence when used in the FTN signaling. Additionally, as the simulation results indicated, the BER performance of the FTN signaling employing the designed optimal pilot sequence showed improvements compared to the FTN signaling employing the conventional Nyquist pilot sequence. Simulation results, at the same SE, showed a significant improvement (about 6 dB) in the MSE of the channel estimation and (about 3.5 dB) BER of FTN signaling employing the designed optimal pilot sequence compared to the one introduced in \cite{ishihara2017iterative} over frequency-selective channels. 

\bibliographystyle{IEEEtran}
\bibliography{IEEEabrv,references}

\end{document}